\documentclass[seceq,preprint]{article}
\usepackage{graphicx}

\begin{document}


\centerline{Rev. Mex. F\'{\i}s. {\bf 51}(3) 316-319 (June 2005)}

\bigskip

\centerline{
\Large Riccati Nonhermiticity with Application to the Morse
Potential }

\bigskip

\centerline{       
Octavio \textsc{Cornejo-P\'erez}, Rom\'an \textsc{L\'opez-Sandoval},
Haret C. \textsc{Rosu\footnote{E-mail: hcr@ipicyt.edu.mx $\qquad$
$\qquad$ quant-ph/0502074 \hfill 4/2005}}
}

\begin{center}
Potosinian Institute of Science and Technology,\\
Apdo Postal 3-74 Tangamanga, 78231 San Luis Potos\'{\i}, Mexico\\


\end{center}

\bigskip



{\small


\noindent
A supersymmetric one-dimensional matrix procedure similar to relationships of the same type between Dirac and Schr\"odinger equations in particle
physics is described at the general level. By this means we are able to introduce a nonhermitic Hamiltonian having the imaginary part proportional to the solution of a Riccati
equation of the Witten type. The procedure is applied to the exactly solvable Morse potential introducing in this way the corresponding nonhermitic Morse problem.
A possible application is to molecular diffraction in evanescent waves over nanostructured surfaces. 

\bigskip


\bigskip

\noindent
Un procedimiento matricial uni-dimensional supersim\'etrico similar a relaciones del mismo tipo entre las ecuaciones de Dirac y Schr\"odinger que usamos recientemente para
el oscilador arm\'onico cl\'asico es presentado de manera concisa en t\'erminos generales. El aspecto nuevo es el uso de par\'ametros constantes por medio de los cuales el Hamiltoniano
se vuelve no herm\'{\i}tico con parte imaginaria proporcional a la
soluci\'on de una ecuaci\'on de Riccati  de tipo Witten, que es caracter\'{\i}stica para el m\'etodo supersim\'etrico. El factor de proporcionalidad contiene los par\'ametros mencionados.
Aplicamos esta t\'ecnica algebraica al oscilador cu\'antico no arm\'onico de Morse obteniendo una forma no herm\'{\i}tica.
Una posible aplicaci\'on es a la difracci\'on molecular en ondas evanescentes sobre superficies nanoestructuradas.  

\bigskip

\noindent {\em Descriptores}: No hermiticidad; supersimetr\'{\i}a;
potencial de Morse

\bigskip


\markboth{    
Cornejo-P\'erez, L\'opez-Sandoval, Rosu
}{            
Morse potential with susy nonhermiticity
}

\newpage


\section{Introduction}

\noindent
We have recently elaborated on an interesting way of introducing imaginary parts (nonhermiticities) in second order differential equations starting from
a Dirac-like matrix equation \cite{A1,A2}. The procedure is a complex extension of the known supersymmetric connection between the Dirac matrix equation and the Schr\"odinger equation \cite{cooper}.
A detailed discussion of the Dirac equation in the supersymmetric approach has been provided by Cooper {\em et al} in 1988,
who showed that the Dirac equation with a Lorentz scalar potential is associated with a susy pair of Schr\"odinger Hamiltonians.
In the supersymmetric approach one uses the fact that the Dirac potential, that we denote by $R$,  is the solution of a Riccati equation with the free term related to the potential function $U$ in the second order linear differential equations of the Schr\"odinger type.

Indeed, writing the one-dimensional Dirac equation in the form
\begin{equation}\label{Dparticulas}
[\alpha p +\beta m +\beta R(x)] \psi (x) = E \psi (x)
\end{equation}
where $c=\hbar =1$, $p=-id/dx$, $m$ ($>0$) is the fermion mass, and $R(x)$ is a Lorentz scalar function representing the potential in which the relativistic particle moves. The wavefunction $\psi$ is a two-component spinor $\big(\begin{array}{c} \psi _1\\ \psi _2 \end{array}\big)$
and the $\alpha$ and $\beta$ matrices are the following Pauli matrices
$$
\sigma _y=
\left( \begin{array}{cc}
0 & - i \\
 i & 0\end{array} \right ) \qquad
{\rm  and}  \qquad 
\sigma _x=\left( \begin{array}{cc}
0 & 1\\
1 & 0 \end{array} \right )~,
$$
respectively.
Writing the matrix Dirac equation in coupled system form leads to
\begin{equation}\label{cs1}
\big[D_x+m+R\big]\psi _1=E\psi _2
\end{equation}
\begin{equation}\label{cs2}
\big[-D_x+m+R\big]\psi _2=E\psi _1~.
\end{equation}
By decoupling, one gets two Schr\"odinger equations for each spinor component, respectively
\begin{equation}\label{cs3}
H_i\psi _i\equiv\big[-D_{x}^{2}+U_i\big]\psi _i =\epsilon \psi _i~, \qquad \epsilon =E^2-m^2~,
\end{equation}
where the subindex $i=1, 2$, and
$$
U_i(x)=\left((m+R)^2-m^2\mp dR/dx\right)~.
$$
One can also write factorizing operators for Eqs.~(\ref{cs3})
\begin{equation}\label{cs4}
A^{\pm} =\pm D_x +m+R
\end{equation}
such that
\begin{equation}\label{cs5}
H_1=A^{-}A^{+}-m^2~, \quad H_2=A^{+}A^{-}-m^2~.
\end{equation}

However, we have employed a so-called complex extension of the method by which we mean that we considered the Dirac potential $R$ as a purely imaginary quantity implying that the Schr\"odinger potentials
$U_i$ are complex, and, as such, we deal with nonhermitic problems. We considered previously the cases of the classical harmonic oscillator and Friedmann-Robertson-Walker barotropic cosmologies, which correspond to the very specific situation in which the Dirac mass parameter that we denoted by ${\rm K}$ was  treated as a free parameter {\em equal} to the Dirac eigenvalue parameter $E$. This is equivalent to Schr\"odinger equations at zero energy, $\epsilon =0$. On the other hand, it is interesting to see how the method works for negative energies, i.e., for a bound spectrum in quantum mechanics. In this paper, we first briefly describe the method and next apply it to the case of Morse potential obtaining a nonhermitic version of this exactly-solvable quantum problem.


\section{Complex extension with a single {\rm K} parameter}
We consider the slightly different Dirac-like equation with respect to Eq.~(\ref{Dparticulas})
\begin{equation} \label{HDM}
\hat{\cal D}_{{\rm K}}W\equiv  [\sigma _y D_{x}+\sigma _x (iR +{\rm K})]W={\rm K}W~,
\end{equation}
where K is a (not necessarily positive) real constant. In the left hand side of the equation, $\rm K$ stands as a mass parameter of the Dirac spinor, whereas
on the right hand side it corresponds to the energy parameter. $R$ is an arbitrary solution of the Riccati equation of the Witten type \cite{w81}
\begin{equation}\label{ricric}
R'\pm R^2=u~,
\end{equation}
where $u$ is the real part of the nonhermitic potential in the Schr\"odinger equations we get.
Thus, we have an equation equivalent to a Dirac equation for a spinor $W=\left(  \begin{array}{cc}
\phi _1\\
\phi _0\end{array} \right )\equiv\left( {\rm \begin{array}{cc}
{\rm w_f}\\
{\rm w_b}\end{array}} \right )
$
of mass $\rm K$ at the fixed energy $E={\rm K}$ but in a purely imaginary potential (optical lattices). This equation
can be written as the following system of coupled equations
\begin{equation}\label{D1}
 iD_{x}\phi _1+(iR+{\rm K})\phi _1={\rm K}\phi _0
\end{equation}
\begin{equation}\label{D2}
 -iD_{x}\phi _0+(iR+{\rm K})\phi _0={\rm K}\phi  _1~.
\end{equation}

The decoupling of these two equations can be achieved by applying the operator in Eq.~(\ref{D2}) to Eq.~(\ref{D1}) . For the fermionic spinor
component one gets
\begin{equation} \label{comp1}
 D^{2}_{x}\phi_1-\Big[R^2-D_x R-i\,2{\rm K}R\Big] \phi_1=0  
\end{equation}
whereas the bosonic component fulfills
\begin{equation} \label{comp2}
 D^{2}_{x}\phi_0-\Big[R^2+D_x R-i\,2{\rm K}R\Big] \phi _0=0  ~.
\end{equation}
This is a very simple mathematical scheme for introducing a special type of nonhermiticity directly proportional to the Riccati solution.
The factorization operators can be written in this case in the form
\begin{equation}\label{comp3}
A^{\pm} =\pm i D_x +{\rm K}+iR
\end{equation}
that allows to write the fermionic equation (\ref{comp1}) as $H_1\phi _1\equiv(A^{-}A^{+}-K^2)\phi _1=0$ and the bosonic equation (\ref{comp2}) as $H_2\phi _0\equiv(A^{+}A^{-}-K^2)\phi _0=0$.



%

\section{Complex extension with parameters {\rm K} and {\rm K'}.}

\noindent
A more general case in this scheme is to consider the following matrix Dirac-like equation
$$
\Bigg[\left( \begin{array}{cc}
0 & - i \\
i & 0\end{array} \right )D_{\rm x}+\left( \begin{array}{cc}
0 & 1\\
1 & 0 \end{array} \right )\left( \begin{array}{cc}
  iR +{\rm K}& 0\\
0 &  iR+{\rm K}\end{array} \right )\Bigg]\left( \begin{array}{cc}
{\rm w}_1\\
{\rm w}_2 \end{array} \right )=
$$
\begin{equation} \label{Dg}
\left( \begin{array}{cc}
{\rm K^{'}}& 0\\
0 &{\rm  K^{'}}\end{array} \right )\left( \begin{array}{cc}
{\rm w_1}\\
{\rm w_2} \end{array} \right )~.
\end{equation}
The system of coupled first-order differential equations will be now
\begin{eqnarray}
\Big[- i D_{\rm x}+ iR+{\rm K}\Big]{\rm w_2}={\rm K^{'}}{\rm w_1}\\ 
\Big[ i D_{\rm x}+ iR+{\rm K}\Big]{\rm w_1}={\rm K^{'}}{\rm w_2}      
\end{eqnarray}

and the equivalent second-order differential equations
\begin{equation} \label{Schrgb}
{ D_{\rm x}}^{2}{\rm w} _{i}
+\Big[\pm  D_{\rm x}R+2 i {\rm K} R+({\rm K^2-K^{'2}})
 - R^2\Big]{\rm w} _{i}=0~,
\end{equation}
where the subindex $i=1,2$ refers to the fermionic and bosonic components, respectively.

Again, introducing the same factorization operators as for the single $\rm K$ case, i.e., $A^{\pm}=\pm iD_x+{\rm K} +iR$, one can write Eq.~(\ref{Schrgb}) in the Schr\"odinger-like form
\begin{equation}\label{schf1}
H_1{\rm w} _1\equiv(A^{-}A^{+}-K^2){\rm w} _1=({\rm K^{'2}-K^2}){\rm w}_1
\end{equation}
and
\begin{equation}\label{schf2}
H_2{\rm w} _2\equiv(A^{+}A^{-}-K^2){\rm w} _2=({\rm K^{'2}-K^2}){\rm w}_2~,
\end{equation}
for the fermionic and bosonic components, respectively. These forms are useful for quantum mechanical applications; see the next section for one of them.

\medskip

\section{Application to the Morse potential}

This potential is frequently used in molecular physics in connection with the dissociation and vibrational spectra of diatomic molecules.
In this case, the Riccati solution is of the type
\begin{equation}\label{RicMorse}
R(x)=A-Be^{-ax}~,
\end{equation}

Therefore, the second-order fermionic differential equation will be
\begin{eqnarray}
D^{2}_{x}{\rm w}_{1}&+&
\left[-\left(\bar{B}\textrm{e}^{-2ax}-\bar{C}_{1}\textrm{e}^{-ax}\right)\right.
+ (K^{2}-K^{\prime 2}) - A^{2} \nonumber\\
&+& \left. 2iK(A-B\textrm{e}^{-ax}) \right]{\rm w}_{1}=0
\end{eqnarray}
where $\bar{B}=B^{2}$, and $\bar{C}_{1}=B(2A+a)$.

The solution is expressed as a superposition of Whittaker functions
\begin{equation}
{\rm w_{1}}=\alpha_{1}\textrm{e}^{ax/2}
M_{\kappa _1, \mu}
\left(\frac{2B}{a}\,\textrm{e}^{-ax}\right)
+\beta_{1}\textrm{e}^{ax/2}
W_{\kappa _1, \mu}
\left(\frac{2B}{a}\,\textrm{e}^{-ax}\right)
\end{equation}
$\kappa _1= \frac{A}{2a}\left(2+\frac{a}{A}-i\frac{2K}{A}\right)$ and $\mu = \frac{A}{a}\left(\frac{K^{'2}-K^2}{A^2}-i\frac{2K}{A}\right)^{1/2}$.

The bosonic equation reads
\begin{eqnarray}
D^{2}_{x}{\rm w}_{2}&+&
\left[-\left(\bar{B}\textrm{e}^{-2ax}-\bar{C}_{2}\textrm{e}^{-ax}\right)\right.
+ (K^{2}-K^{\prime 2}) - A^{2} \nonumber\\
&+& \left. 2iK(A-B\textrm{e}^{-ax}) \right]{\rm w}_{2}=0
\end{eqnarray}
where $\bar{B}=B^{2}$, and $\bar{C}_{2}=B(2A-a)$.

The solution is a superposition of the following Whittaker functions
\begin{equation} \label{superposition}
{\rm w_{2}}=\alpha_{2}\textrm{e}^{ax/2}
M_{\kappa _2, \mu}
\left(\frac{2B}{a}\,\textrm{e}^{-ax}\right)
+\beta_{2}\textrm{e}^{ax/2}
W_{\kappa _2 , \mu}
\left(\frac{2B}{a}\,\textrm{e}^{-ax}\right)
\end{equation}
where $\kappa _2=\frac{A}{2a}\left(2-\frac{a}{A}-i\frac{2K}{A}\right)$ and the $\mu$ subindex is unchanged.

If we now place ourselves within the quantum mechanical (hermitic) Morse problem we should take $\beta _2=0$ and $K=0$ in order to achieve the exact correspondence with the bound spectrum problem and
eliminate the non-hermiticity.
Moreover, the following well-known connection with the associated Laguerre polynomials
\begin{equation} \label{Laguerre1}
M_{\frac{p}{2}+n+\frac{1}{2}, \frac{p}{2}}(y)= y^{\frac{p+1}{2}}e^{-y/2}L_{n}^{p}(y)~, \quad y=\frac{2B}{a}e^{-ax}
\end{equation}
can be used in our case with the following identifications
$$
\frac{p}{2}=\frac{K'}{a}~, \qquad
K'=(A-an)
$$
i.e.,
$$
p=2\left(\frac{A}{a}-n\right)~.
$$
Then we can write the solution of the hermitic bosonic problem in the well-known form
\begin{equation}\label{wn}
{\rm w} _{2,n}(y) =\alpha _2\left(\frac{2B}{a}\right)^{\frac{1}{2}}y^{\frac{A}{a}-n}e^{-y/2}L_{n}^{2(\frac{A}{a}-n)}(y)~.
\end{equation}
If we want to approach the nonhermitic problem we define by analogy with Eq.~(\ref{Laguerre1})
\begin{equation}\label{Laguerre2}
M_{\kappa _2,\mu}(y)= y^{\mu +\frac{1}{2}}e^{-y/2}L_{\kappa _2-\mu -\frac{1}{2}}^{2\mu}(y)~,
\end{equation}
where $\kappa _2$ and $\mu$ are the complex parameters mentioned before and the symbol corresponding to the associated Laguerre polynomial representing now a Laguerre-like function introduced
by definition through Eq.~(\ref{Laguerre2}).
The wavefunction of the nonhermitic problem can be written as follows
\begin{equation}\label{wnonherm}
{\rm w} _{2,nonherm}(y) =\alpha _2\left(\frac{2B}{a}\right)^{\frac{1}{2}}y^{\mu}e^{-y/2}L_{\kappa _2-\mu -\frac{1}{2}}^{2\mu}(y)~.
\end{equation}

In the case of the nonhermitic fermionic problem, the formulas are similar with the replacement of $\kappa _2$ by $\kappa _1$. Thus:
\begin{equation}\label{wnonherm1}
{\rm w} _{1,nonherm}(y) =\alpha _1\left(\frac{2B}{a}\right)^{\frac{1}{2}}y^{\mu}e^{-y/2}L_{\kappa _1-\mu -\frac{1}{2}}^{2\mu}(y)~.
\end{equation}

\medskip

In conclusion, the supersymmetric connection between Dirac-like equations and Schr\"odinger equations, in a simple complex extension form, has been applied here in the quantum context
of the Morse potential. However, in the pure quantum case, the results of this note seem to be only of mathematical interest. A natural question is how different types of nonhermiticities can be engineered.
As we noticed in our previous research \cite{A1}, physical optics is closer to real applications. In particular, one can think to the diffraction of diatomic molecules in evanescent fields because such fields have imaginary wavenumbers and we know that Schr\"odinger equations are similar to
Helmholtz equation in the paraxial approximation.
A specific experimental setup could be very similar to that discussed recently
by L\'ev\^eque and collaborators \cite{lev02} in their study of diffractive scattering of cold atoms from an evanescent field, spatially modulated by an array of nanometric objects with high
index of refraction and subwavelength periodicity deposited on a glass surface. The evanescent wavefield is created by a totally internally reflected laser beam and is strongly modulated by the configuration
of the nanostructure. The calculations are not easy as one should tackle Helmholtz equations in complicated geometries. The task is to obtain the configuration of the nanostructure that is able to produce
the evanescent field corresponding to the nonhermitic part of our Morse problem. This is experimentally only a challenging possibility for the time being.
\medskip
\medskip




%

\bigskip
\bigskip

\noindent

The four figures next have not been included in the RMF published
version.


\newpage





\begin{figure}  [x] 
\centerline{
\includegraphics[scale=1]{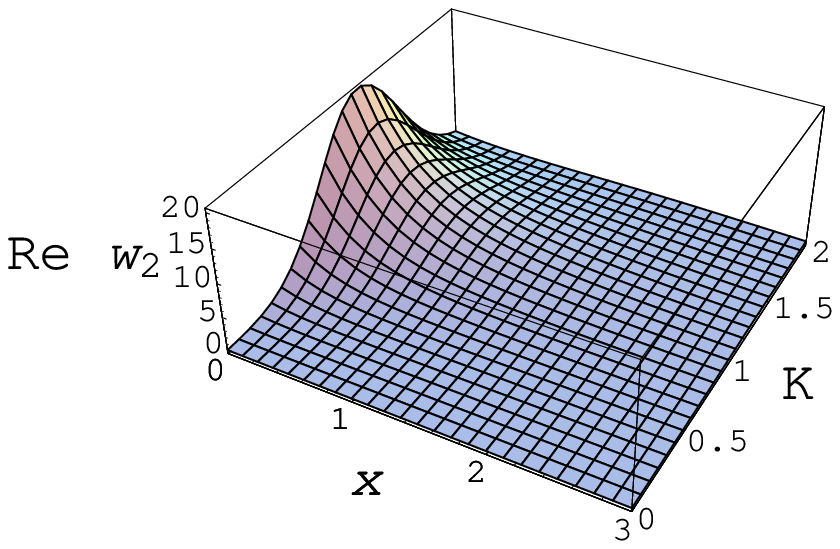}}
\caption{The real part of the bosonic mode ${\rm w}_{2, nonherm}(y)$ for ${\rm x}\in [0,3]$ and the nonhermiticity parameter in the range ${\rm K\in[0,2]}$.
The rest of the parameters have the following values: $A=1$, $B=2$, $a=0.5$, $K'=2$, and $\alpha _{1,2}=1$ in all the figures of this paper.
} \label{figmorse1}
\end{figure}

\begin{figure}   [x]  
\centerline{
\includegraphics[scale=1]{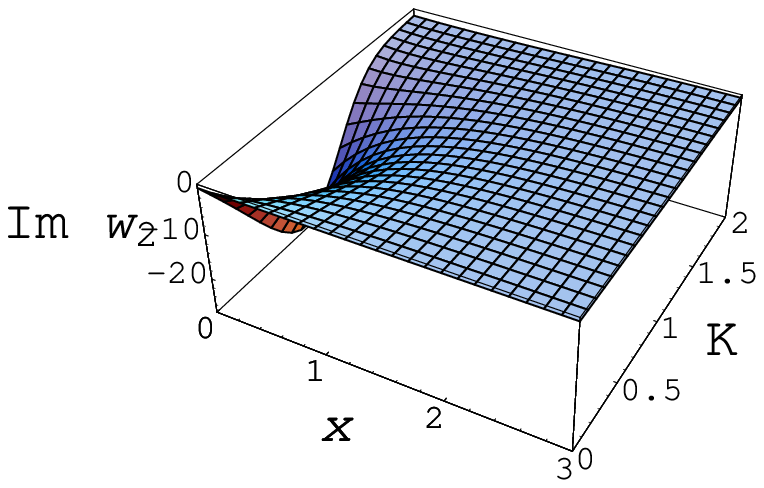 }}
\caption{The imaginary part of the bosonic mode ${\rm w}_{2, nonherm}(y)$ for the same range of ${\rm x}$ and ${\rm K}$.
} \label{figmorse1}
\end{figure}

\begin{figure}  [x]   
\centerline{
\includegraphics[scale=1.1]{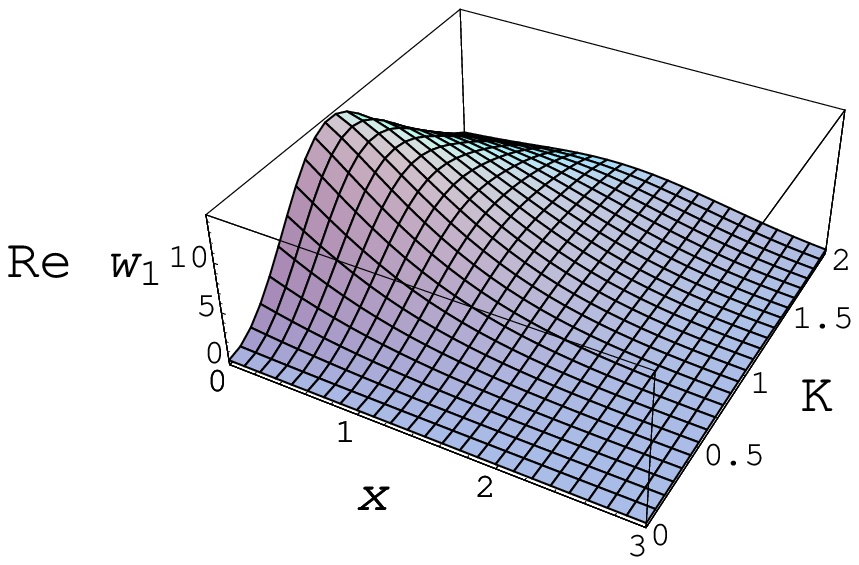 }}
\caption{The real part of the fermionic mode ${\rm w}_{1, nonherm}(y)$ for ${\rm x}\in [0,3]$ and the nonhermiticity parameter in the range ${\rm K\in[0,2]}$.
} \label{figmorse1}
\end{figure}

\begin{figure}  [x]  
\centerline{
\includegraphics[scale=1.1]{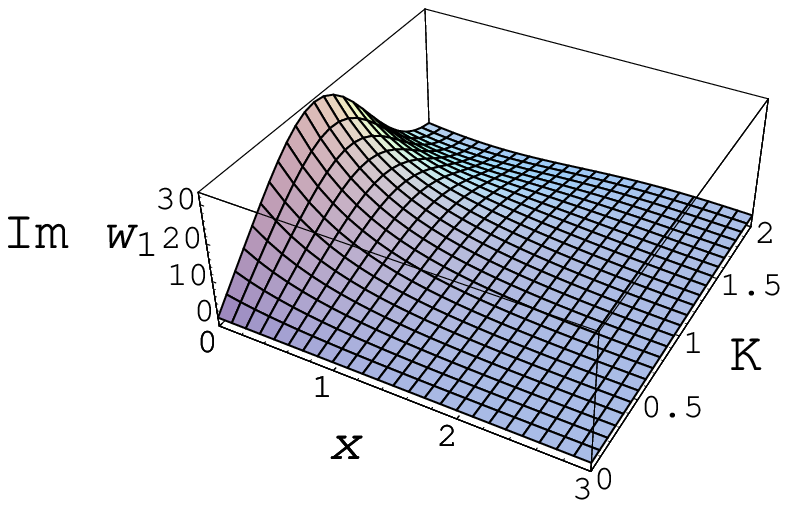 }}
\caption{The imaginary part of the fermionic mode ${\rm w}_{1, nonherm}(y)$ for ${\rm x}\in [0,3]$ and ${\rm K\in[0,2]}$.
} \label{figmorse1}
\end{figure}

\end{document}